\newcommand{\be}{\begin{equation}}
\newcommand{\ee}{\end{equation}}

\documentclass{article}
\pdfoutput=1
\usepackage{amsmath,mathtools}
\usepackage{amsfonts}
\usepackage{braket}
\usepackage{graphicx}
\usepackage[font={scriptsize,it}]{caption}
\usepackage{amssymb}

\begin{document}
\thispagestyle{plain}
\begin{center}
 \LARGE
A perturbative expansion for entanglement entropy in string theory
\vspace{0.7cm}
\large 

\textsc{Andrea Prudenziati}
\vspace{0.4cm}

\textsl{International Institute of Physics,\\ Universidade Federal do Rio Grande do Norte, \\Campus Universitario, Lagoa Nova, Natal-RN 59078-970, Brazil}
\vspace{0.9cm}
\end{center}

\begin{abstract}
\noindent  
We derive a perturbative expansion for space time entanglement entropy in string theory by comparing replica trick constructions on the target space and on the worldsheet. Requiring the two approaches to match implies a set of constrains on the amplitudes involved and produces a non-trivial relation among target space and world sheet entanglement. We will be mainly interested in bosonic string on a flat background  but also discuss the superstring generalization.
\end{abstract}


\section{Introduction}

Entanglement entropy in conformal field theories is a well developed subject where problems can be approached by purely quantum field theory techniques, as for instance the celebrated replica trick construction, \cite{Calabrese:2004eu} and \cite{Holzhey:1994we}, or holographic methods as the Ryu-Takayanagi formula, \cite{Hubeny:2007xt} and \cite{Ryu:2006bv}. Little however is known when string theory is involved, mainly because the extended nature of strings hinders an exact formulation of the problem and a strict approach would need to consider second quantization. It is nonetheless clear that understanding how to compute string theory entanglement entropy for a certain spatial region would be a valuable tool in many contexts, the main example perhaps being black holes.  Different ideas have been explored in the past: \cite{He:2014gva} studied an analytic continuation of the space-time replica trick manifold in $1/n$ leading to well understood orbifold backgrounds ( see also \cite{Dabholkar:1994ai} and \cite{Dabholkar:1994gg}). \cite{Donnelly:2016jet} instead considered a specific setting by choosing the closed string theory dual to two dimensional Yang-Mills while \cite{Balasubramanian:2018axm} pursued the path of a string field theoretic approach. In \cite{Prudenziati:2016dbc} we instead focused on the pure world sheet problem by considering contributions from winding states and only on a second stage its target space interpretation. Other related papers are \cite{Hartnoll:2015fca}, \cite{Huang:2017lbl}, \cite{Lee:2018gyq}, \cite{Niarchos:2017cdz} and \cite{Quijada:2017zif}.

In the present paper we will work our way around the technical obstacles by comparing two different approaches for a perturbative formulation. On one side we will work under the assumption that the target-space entanglement in string theory should ultimately be generated by world sheet degrees of freedom; we will then construct an ansatz to compute entanglement entropy for a certain target space region by studying \emph{world sheet} entanglement for the corresponding world sheet region, induced by the inverse embedding\footnote{the obvious issue that the embedding is itself determined by the very degrees of freedom for which we are considering the entanglement will be discussed}. Also, exploiting the free nature of the string theory Lagrangian, the twist operators involved will be seen to coincide with orbifold twisted states, so that the problem is translated perturbatively to a computation of certain string correlation functions of orbifold twist operators on Riemann surfaces. On the other side the $1/n$ analytic continuation of the target space replica trick manifold also leads to a world sheet perturbative expansion on orbifold background where, as we will see, an arbitrary number of intermediate twisted states insertions localized at the cut should be included. The requirement that these two results should agree then implies specific conditions on both sides, that will characterize the form of the perturbative series of correlation functions. 

In sections \ref{section2} and \ref{section3} we present the two complementary constructions for a perturbative series from a world sheet and target space perspective, fixing the general requirements. In section \ref{section4} we ask for both set of amplitudes to agree, leading to additional conditions on the two sides. In section \ref{section5} we discuss how to treat ghosts and solve the problem regarding the correct ordering of gauge fixing of the world sheet metric and replica construction. In section \ref{section6} we present the complete formulas and derive the first perturbative terms while in section \ref{4psphere} we use this result to derive an explicit expression for the first non trivial contribution to the entanglement entropy characterizing our proposal. Finally in section \ref{section7} we show how to generalize, at least formally, for fermions by deriving a corresponding twist operator computation. We then present our conclusions, open problems and possible applications.

\section{Target space entanglement from twisted states correlators}\label{section2}

We consider for simplicity flat space in d-dimensional $\mathbb{R}^d$, parametrized by coordinates
$x^0,x^1, \cdots,x^{d-1}$. The entangling region $A$ lives on the space-like slice $x^0=0$ and is defined by the inequality $x^1\geq 0$. We would like to compute the corresponding entanglement entropy for some string theory; in this section we will consider bosonic string theory ( for which $d=26$ even though we will not make the substitution explicit), and later on we will consider superstrings where additional interesting issues arise. Moreover we will concentrate just on the scalar fields $X^i$ forgetting for the moment the world sheet metric; we will come back to it.

\begin{figure}[h]
\centering
\includegraphics[width=0.6\textwidth]{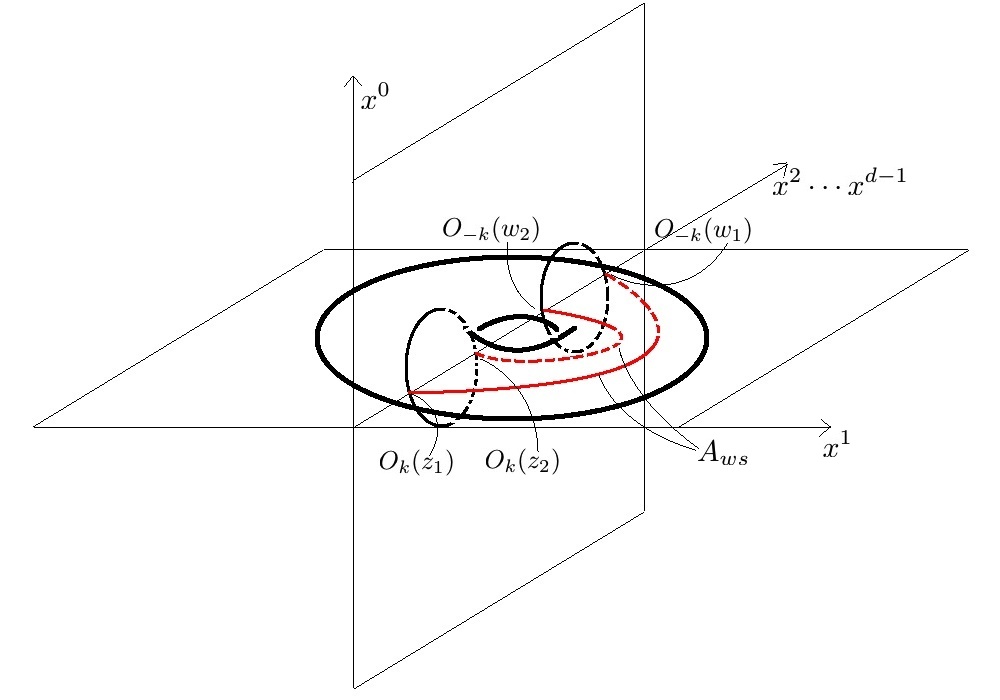}
\caption{A genus one world sheet intersects the plane $x^1=0$ in the two black circles drawn in the figure above, and the $x^0$ plane inside the $x^1\geq 0$ region in the two red arcs that represent then $A_{ws}$. The boundary $\partial A_{ws}$ is given by the four points where the four twist-antitwist operators are inserted. }\label{figuraa}
\end{figure}

To start with let us consider a two dimensional sigma model with fixed embedding $X^i(z,\bar{z})$ of the world sheet into the target space, plus an additional field $\phi(z,\bar{z})$ that represents our local degrees of freedom. We can define the \emph{target space} entanglement entropy that "comes from $\phi$" as the target space entanglement associated to the Hilbert space of states produced by generic world sheet operators constructed out of the $\phi$ field and its derivatives. We could then argue that the contribution from that specific embedding  could be computed by the \emph{world sheet} entanglement entropy for the induced  region $A_{ws}$, that we can formally represent as $A_{ws}=X^{-1}(A)$ ( for $X^{-1}$ we mean the inverse map from the target to the world sheet); note that in general $A_{ws}$ can be made of an arbitrary number of disconnected sets, even if $A$ is a connected region. Moreover, as $\partial A_{ws}$ is made by a set of points on the world sheet, determined by the intersection of the two hyperplanes $x^0=x^1=0$ with the embedding $X^i(z,\bar{z})$, the replica trick construction is  implemented by considering appropriate twist and anti twist operators located at these points and acting as permutations on the $n$ copies of $\phi$. Figure \ref{figuraa} should make the intuitive idea behind this formulation clear.

The obvious problem when considering string theory is that some of the $\phi$ coincides with the embedding maps themselves, so that the region $A_{ws}=X^{-1}(A)$ will not be constant under  path integration! We can visualize this by first fixing a reference embedding $X_{ref}$ and then integrating  over a small perturbation $\Delta X=X-X_{ref}$. The equalities $X^0_{ref}=X^1_{ref}=0$ determine the world sheet points that are the boundary of $A_{ws}$ at $\Delta X=0$, and their change of location on the Riemann surface under integration of $\Delta X$ will come from 
\be\label{uno}
\Delta (\partial A_{ws})\sim\Delta X^{-1}|_{X^0_{ref}=X^1_{ref}=0}.
\ee
As there is not any clear physical motivation for keeping $\partial A_{ws}$ fixed by requiring $\Delta (\partial A_{ws})=0$ on path integral configurations, the alternative we will consider is a milder construction whose rationale will be fully understood only later on. 

The ansatz for a perturbative world sheet computation of target space entanglement entropy for $A$ then passes through the computation of the Renyi entropy partition function  $Z_n(A_{ws})$; this will contain generic correlation functions constructed as follows:
\begin{itemize}
\item Consider a Riemann surfaces with generic $l$ integrated insertions of twist operators $O_{tw}$ and an equal number of anti twist operators $O_{tw}^{-1}$. These choices will cover any possible intersection of the world sheet embedding with the hyperplanes $x^0=x^1=0$ \footnote{The case of an hyperplane being tangent to the world sheet embedding is obtained by colliding one twist and anti twist operators whose OPE is just the identity.}.
\item Fix the center of mass of the string fields $X_0^0,X_0^1$ at the twist and anti twist operator positions, and the momentum $p_0^0,p_0^1$ to zero, and integrate only over the oscillations. Roughly speaking we want the string to be localized at the intersection with $A$ but still allow quantum oscillations nearby to produce entanglement. 
\item Compute the amplitude and sum over the integer $l$ from zero to infinity\footnote{some amplitudes will possibly vanish depending on the topology}
\end{itemize}
Hiding for the moment the ghost sector the result is:
\begin{align}\label{propw}
&\log Z_{n}(A_{ws})= \\
&=\sum_{g}\sum_{l=0}^{\infty}\int_{\Sigma_g}dz_1\cdots dw_l<O_{tw}(z_1)O_{tw}^{-1}(w_1)\cdots O_{tw}(z_l)O_{tw}^{-1}(w_l)>_{\Sigma_g}|_{\substack{\text{$X_0^0=x^0=0$} \\ \text{$X_0^1=x^1=0$} \nonumber \\  \text{$p_0^0=p_0^1=0$}}}
\end{align}
Some concerns with this methodology are obvious, in particular on the interpretation of the result as space time entanglement entropy; furthermore open questions remain as, for instance, if all the stringy degrees of freedom should evenly contribute to the target space entanglement ( or reformulating, if we should include twist operator acting on all the world sheet fields). Thus it is important to stress that equation (\ref{propw}) should be intended mainly as a generic ansatz, and the discussion leading to it as the intuitive physical interpretation of what these amplitudes may actually compute. Only when we will compare this result with the target space approach we will finally put all this hand waving argument on firmer ground. 

\section{Twisted operators and orbifolds}\label{section3}

It is well known that a two dimensional path integral over the replica trick $n$-sheeted manifold can be reduced to a single sheet computation for $n$ fields $X_a$, $a=1,\cdots,n$. This is achieved by placing appropriate couples of twist and anti-twist operators at the boundary of the entangling surface so that crossing the  branch cut  that connects the operator positions within each couple permutes the fields as $X_a\rightarrow X_{a+1}$ and $X_n\rightarrow X_{1}$. If the theory is free, as in the case of the bosonic string Lagrangian on flat space, we can construct linear combinations of fields ( whose Lagrangian will still be free) as:
\be\label{due}
X_k=\sum_{a=1}^{n} X_a e^{-i 2 \pi a \frac{k}{n}}   \;\;\;\;\;\; k=0,\cdots, n-1.
\ee
The field $X$ has an additional index for its target space dimension and from now on we will keep using an upper index for this  and a lower one for the sheet number ( or $k$-linear combination). We can then construct  complex couples as $X=X^0+i X^1$ and $\bar{X}=X^0-i X^1$, so that (\ref{due}) will be applied to the first of these complex fields \footnote{ or any other $X_a^{i}=X_a^{2(i-1)}+i X_a^{2i-1}$ with $i=2,\cdots,d/2$, being by convention $X^1 \equiv X$. We will not use the real fields any more so that hopefully no confusion will arise by naming $X$ both the real and complex scalars.} and its complex conjugate reads
\begin{equation}\label{due2}
\bar{X}_k=\sum_{a=1}^{n} \bar{X}_a e^{i 2 \pi a \frac{k}{n}}.
\end{equation}
 The advantage of this $k$-basis is that turning around a twist operator will now produce just a phase 
\be \label{perso}
X_k\rightarrow X_{k}e^{i 2 \pi \frac{k}{n}} \;\;\;\;\; \bar{X}_k\rightarrow \bar{X}_{k}e^{-i 2 \pi \frac{k}{n}}
\ee 
and the opposite for an anti twist. The operator that does so for the fields $X_k, \bar{X}_k$ ( and acts as the identity for $l\neq k$ and any other field) is named $O_k$. The replica trick computation is then reproduced by operator insertions at the boundary of the entangling region 
( $\partial A_{ws}=\{z_1,w_1,z_2,w_2,\cdots, z_l,w_l\}$) as:
\be\label{tre}
\prod_{k=0}^{n-1}< O_k(z_1)O_{-k}(w_1)\cdots O_k(z_l)O_{-k}(w_l)>.
\ee 
We then have an additional condition to include in the world sheet construction spelled so far:
\begin{itemize}
\item By changing basis as in (\ref{due})-(\ref{due2}) we can replace the twist-antitwist operator insertions inside the world sheet amplitudes (\ref{propw})  with $l$ couples of twist-anti twist operators in the specific form (\ref{tre}), for $X, \bar{X}$ as well as any other field $X^i, \bar{X}^i$ ( with $i=2,\cdots, d/2$), for which the corresponding twist-anti twist operators are $\prod_{k=0}^{n-1}O_k^i$ and $\prod_{k=0}^{n-1}O_{-k}^i$.
\end{itemize}

A key observation now is that an analogous behaviour to (\ref{perso}) happens for $\mathbb{Z}_n$ orbifolds where the fields $X_k, \bar{X}_k$ are the twisted sectors produced by $O^{orb}_k$. Just by definition we can then identify $O^{orb}_k=O_k$. 

\begin{figure}[h]
\centering
\includegraphics[width=0.6\textwidth]{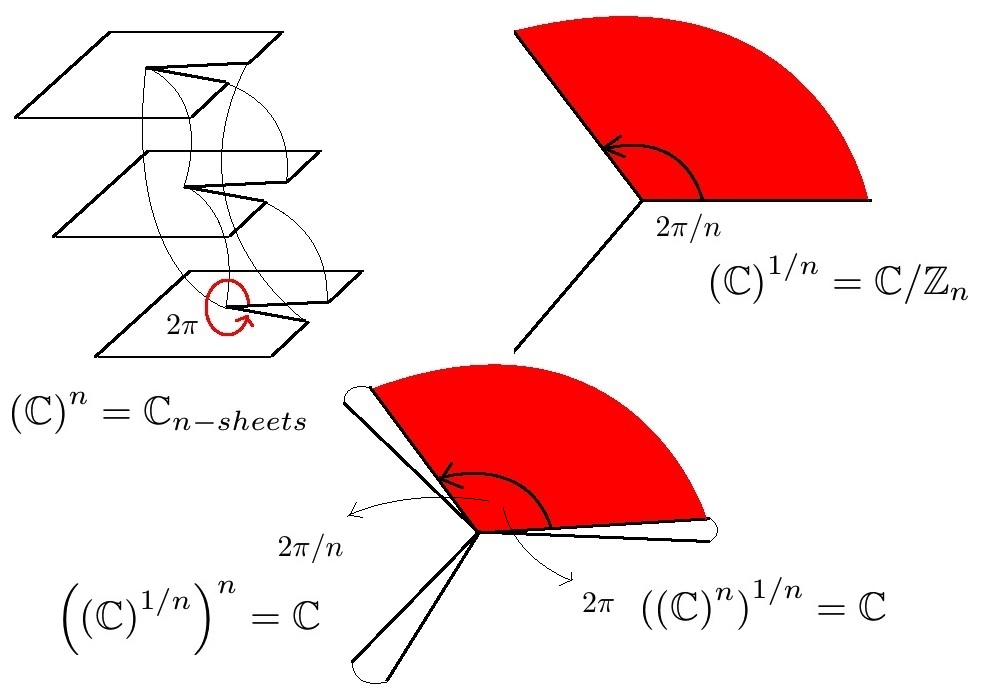}
\caption{A standard replica trick manifold $\mathbb{C}^n$ is represented in the top left picture ( $n=3$ in this case) with cuts identified explicitly. In the top right image we have coloured in red the coset region for the $\mathbb{C}^{1/n}=\mathbb{C}/\mathbb{Z}_n$ orbifold, where $Z_n$ acts rotating points in $\mathbb{C}$ by an angle $2\pi/n$ ( and again $n=3$ here). The final picture represents the $\mathbb{C}$ plane either obtained as $\left(\mathbb{C}^{1/n}\right)^n$  if we start with the orbifold coset in the red region and then construct a replica trick manifold for it, or $\left(\mathbb{C}^n\right)^{1/n}$ if we start with the red region being the $\mathbb{C}$ plane itself and then replicate it and orbifold the replicated manifold. Note that both the replica trick and the orbifold lead to a delta function singularity at the $x^0=x^1=0$ fixed point, but this cancels when the two constructions are applied in series.  }\label{figurab}
\end{figure}

Let us now took a step back and examine the \emph{target} space replica trick construction. The trick is to consider, instead of the standard $n$-sheeted manifold $M^n$,  its "inverse" $M^{1/n}$ defined so that $\left( M^{1/n}\right)^n=\left( M^{n}\right)^{1/n}=M$ ( note that we always mean $n\in\mathbb{N}$). As the goal is to analytically continue in $n$ around $n=1$ we should not be worried about which among $M^n$ and $M^{1/n}$ we construct to derive the Renyi entropy, which is either $Tr \rho_A^n$ or what we may loosely speaking indicate as $Tr \rho_A^{1/n}$, but is really defined as the path integral over  $M^{1/n}$ \footnote{ we may be worried with the convergence of the path integral on $M^{1/n}$ because if we diagonalize $\rho_A$ its eigenvalues $\lambda_i$ are always in between zero and one and $\sum_i\lambda_i=1$ so that $\sum_i\lambda^n_i$ obviously converges but $\sum_i\lambda^{1/n}_i$ may give troubles. However we ultimately know that the orbifold amplitudes are well behaved in string theory so that such issues are in fact not a problem in the cases considered here. }. For $M=\mathbb{C}$ and an half space entangling surface $x^0=0, x^1\geq 0$ the two constructions are schematically represented in figure \ref{figurab} and we can easily see that our $\mathbb{C}^{1/n}$ is just an usual $\mathbb{C}/\mathbb{Z}_n$ orbifold. The "inverse" replica trick construction for the target space $\mathbb{R}^d$  is thus implemented by a path integral computation over a $\mathbb{R}^{d-2}\times \mathbb{C}/\mathbb{Z}_n $ background.
This idea was used in \cite{He:2014gva} together with certain assumptions on the analytic continuation to compute  string one loop amplitudes on orbifold background and obtain perturbative contributions for the target space entanglement entropy. 

Now an important remark: from a space time perspective the replica trick manifold comes from the path integral construction of $Tr(\rho_A)^n$, with $\rho=\ket{0}\bra{0}$. The cuts along $A$ are nothing but fixed field configurations for evaluating matrix elements of $\rho_A$ in the Hilbert space of the theory, and the periodic identification amounts to their multiplication and final trace. If our theory contains additional states  localized nearby the cut and not included into the asymptotic Hilbert space of "in" and "out" states, these should be included as new matrix elements of $\rho_A$. But in the orbifold computation such states do indeed exist! They are the twisted orbifold states localized around the orbifold fixed point(s) (that are just the boundary of the entangling region), and generated by insertion of an arbitrary number of orbifold twist operators. We are then led to the target space proposal for the perturbative computation of the Renyi entropy $Z_{1/n}(A)$:
\begin{itemize}
 \renewcommand{\labelitemi}{\tiny$\blacksquare$}
\item Compute the full perturbative series of generic correlation functions of twist operators on the orbifold background $\mathbb{R}^{d-2}\times\mathbb{C}/\mathbb{Z}_n$, summing over genus and number of integrated twist operator insertions. As $\mathbb{Z}_n$ acts only on $\mathbb{C}$ the twisted states will be just $O_k$ ( that would be $O_k^1$). The condition for the amplitude not to vanish is that the total twist number is $\sum_i k_i\in \mathbb{Z}$, and all such twist operators should a priori be included.
\end{itemize}
Once more hiding the ghost sector the result becomes:
\begin{align}\label{propt}
&\log Z_{1/n}(A)=\nonumber \\
&=\sum_{g}\sum_{l=0}^{\infty}\sum_{\sum_i k_i\in \mathbb{Z}}\int_{\Sigma_g}\hspace{-0.3cm}dz_1\cdots dw_l < O_{k_1}(z_1)O_{k_2}(w_1)\cdots O_{k_{2l-1}}(z_l)O_{k_{2l}}(w_l)>_{\Sigma_g}^{\mathbb{R}^{d-2}\times\mathbb{C}/\mathbb{Z}_n}.
\end{align}
Before continuing let me address a usual concern with string theory on conical spaces; whenever an analytic continuation in $n$ is required the theory is off-shell for $n$ non-integer. In the following we will not try to make sense of the amplitudes for generic real $n$, but we will instead define them as some appropriate analytic continuation of well defined functions of integer $n$. \footnote{ the question of which analytic continuation should be chosen is complicated ( see for example \cite{He:2014gva} for a proposal) an it will not be addressed here.}  

\section{Merging the two constructions}\label{section4}

We are now in place for connecting the two constructions. We argued that, to produce target space entanglement from a world sheet perspective as in (\ref{propw}), we needed to compute Riemann surface amplitudes (\ref{tre}) for any $O_k^i$, involving an even number of integrated twist and anti twist operators at the boundary of $A_{ws}$, with the additional requirement that the center of mass  positions of the nearby strings had to be localized at $X^0=X^1=0$. We have also seen that one of these operators, $O_k^{i=1}= O_k$ have an analogous behaviour to the $\mathbb{C}/\mathbb{Z}_n$ orbifold twisted states operator so that we could identify the two. The nice additional property is that these operators are localized by construction around the orbifold fixed point $X^0=X^1=0$ so that no additional assumption is needed on the path integral for them. Furthermore from the target space perspective the orbifold background itself comes naturally as an "inverse" replica trick manifold and the inclusion of twisted states around the fixed point is indeed required, if we want to evaluate matrix elements of the reduced density matrix $\rho_A$ on the total local Hilbert space ( that indeed contains twisted states). 

Based on the obvious similarities, and ultimately the assumption that entanglement entropies on both sides should agree, we are then led to propose the equivalence between the two constructions. Note that this proposed equivalence is not trivial at all as it identifies the Renyi entropy computed with a world sheet replica trick with $n$-sheets $Z^n(A_{ws})$ to the one obtained from the target space "inverse" replica trick of the $\mathbb{Z}_n$ orbifold and intermediate twisted states $Z^{1/n}(A)$, provided some additional conditions. 

Looking at each side separately these constraints may appear in some sense too restrictive, but the nice feature of this construction is that they have a natural justification from the corresponding complementary picture. The additional assumption for the world sheet construction (\ref{propw})-(\ref{tre}) is:
\begin{itemize}
 \item The twist operator insertions we should include act only on the $X,\bar{X}$ complex fields and not on the other space time dimensions, which is clear as seen from the target space ( as only these twisted states can be included) but an interesting statement from the world sheet point of view \footnote{The inclusion of $O_k^{i\neq1}$ would indeed create some problems also on the world sheet side by fixing all the center of mass and momenta for the coordinates $X^{i\neq1}$ at the operator positions, as these as well are orbifold twist operators acting on different $\mathbb{C}$ planes, so that seen from the target space  the world sheet would appear squeezed at $\partial A_{ws}$ in every direction.}.
 \end{itemize}
For the target space construction (\ref{propt}) instead we should ask:
 \begin{itemize}
 \renewcommand{\labelitemi}{\tiny$\blacksquare$}
\item The generic twisted states amplitudes (\ref{propt}) will be restricted to be of the form (\ref{tre}), that is $l$ couples of twist and anti twist operators all with the same index $k$. This comes from the world sheet free field replica trick construction reviewed at the beginning of this section but is unexpected from the target space perspective where more general cases are permitted (any set of indexes $k_1\cdots k_l$ provided their sum is some integer). 
\end{itemize}

Requiring the above points the two perturbative series agree. An important fact to emphasize is that computing world sheet amplitudes of twist operators automatically requires projection of the spectrum to invariant states (along all the non trivial closed cycles of net twist zero, summarized by condition (\ref{dieci}) in the Appendix) and addition of the twisted sector (the closed strings circling twist operators). This  mimics the twisting construction that orbifold backgrounds impose on amplitudes. Also correlation functions without any operator insertion should be computed  including twisting as they need to be recovered as OPE of twist and antitwist operators inside two point functions.

Before developing more in detail the correspondence let us discuss the ghost sector.

\section{Ghosts}\label{section5}

An important point to understand before moving on is how to treat ghosts. In \cite{Prudenziati:2016dbc} it was noted that the critical issue is to understand if the procedure of gauge fixing the two dimensional metric by the diff $\times$ Weyl  symmetry should be done before or after the $n$-fold construction of the replicated world sheet. 

Results are obviously different as in the first case we would have moduli, and not the metric, as the fields to be replicated on the various sheets. Since a modulus is constant on each sheet, identification through the cut means identification on the full replicated world sheet; so the same moduli set describing the original base Riemann surface apply to the $n$-sheet foliation, identical for each sheet, and a final overall integral. Then we end up with an higher genus surface of a very special kind, described by less moduli than its topology would generally require. 

In the second case we would have a two dimensional metric on each sheet as the field to be replicated, and the gauge fixing occurring at the level of the foliated Riemann surface; then each sheet would end up with its own set of moduli and the replicated world sheet becomes a generic higher genus surface. In \cite{Prudenziati:2016dbc} it was seen that, for the very specific case of one loop open string amplitudes, the mechanism of tadpole cancellation makes the two descriptions equivalent, but for any other topology this is not expected. 

From the world sheet point of view we do not have a clear reason to select either of the two choices, even if the first scenario is the most desirable leading to  much easier computations. But the target space description of entanglement by using orbifold backgrounds comes again in help. As we want to make contact with it, the foliation in the world sheet replica trick construction should be necessarily implemented  by using twist operators on the base Riemann surface, as done so far. But note that this implies that the position of the cut on any sheet (that starts and ends at the twist-anti twist operator insertions) should be the same, thus constraining the higher genus surface obtained by foliation to be of the special type we encountered when gauge fixing is done \emph{before} foliation. And indeed when computing orbifold correlation functions it is well known that the ghost sector is unaffected as twists do not change the world sheet topology. Had we chosen the other way around, the moduli associated to the position of the cut on each sheet would have been generic, and a twist operator description impossible. 

 Some examples we encounter from orbifolds that may help to clarify this point are the follows: the sphere with two couples of twist - anti twist operators ($l=2$) computed in \cite{Dixon:1986qv} has a  $n$-fold cover of genus $g=n-1$ but, instead of the expected $3n-6$  moduli, turns out to be described by a single complex parameter: the position of the fourth operator not fixed by $SL(2,\mathbb{C})$ invariance; analogously for $l$ couples of operators the $n-$fold cover has genus $g=(n-1)(l-1)$ and we would expect $3g-3=3(n-1)(l-1)$ moduli, while the twist operator amplitude only contains $2l-3$ complex numbers, again the unfixed operator positions (see appendix \ref{appa}). Similar higher genus cases are described in \cite{Atick:1987kd}. 
 
\section{Perturbative expansion}\label{section6}

It is time to collect all the information and be explicit on the perturbative expansion needed to compute the entanglement entropy $S(A)$. What we have claimed so far is that, for any $n \in\mathbb{N}$:
\begin{align} \label{sette7}
&\log Z_{1/n}(A)=\log Z_{n}(A_{ws})= \nonumber \\
&=\sum_{g}\sum_{l=0}^{\infty}\sum_{k=0}^{n-1}\int_{\Sigma_g}dz_1\cdots dw_l< O_k(z_1)O_{-k}(w_1)\cdots O_k(z_l)O_{-k}(w_l)>_{\Sigma_g}^{\mathbb{C}/\mathbb{Z}_n}.
\end{align}
These amplitudes also involve an integral over the relevant moduli space of $\Sigma_g$ with $2l$ insertions, with the usual measure being the contraction of the $b$ ghost with Beltrami differentials and, for genus zero and one, an appropriate number of $c$ ghost insertions. The above formula directly implies:
\be\label{ff}
S(A)=\lim_{n\to 1_{\pm}} \frac{n}{n-1}\log \left(\frac{Z_{1/n}(A)}{Z^{1/n}}\right)=\lim_{n\to 1_{\mp}} \frac{1}{1-n}\log \left(\frac{Z_n(A_{ws})}{Z^n}\right)
\ee
where $1_{\pm}=1\pm\epsilon, \;\; 1\gg\epsilon>0$ \footnote{Or alternatively there is a sign difference in between the target space and worldsheet entanglement entropies.}. 

 Weighting the amplitudes in (\ref{sette7}) with a string coupling constant factor of $\lambda^{2g-2+2l}$, we can obtain the order by order expansion of $S(A)$ as:
\be \label{sette8}
S(A)=\lim_{n\to 1}\frac{n}{n-1}(T_n^2 -\frac{1}{n} T_1^2+ S_n^2[O_k,O_{-k},I ]+
\ee
\[
+ \lambda^2\left(S_n^2[O_k,O_{-k},O_k,O_{-k}]+T_n^2[O_k,O_{-k}]+G_n^2-\frac{1}{n}G_1^2\right) + O(\lambda^4)).
\]
A few remarks on this result. First of all $S^2, T^2, G^2$ are the sphere, torus and genus two amplitudes with the dependence by $n$ explicitly shown. The operator insertions are included inside brackets "$[\hspace{0.2cm}]$" and the sum over $k$ as well as the integrals other the positions are implicit. Some amplitudes like the sphere with a single twist-anti twist couple would vanish because of the c-ghost zero modes integral, so the identity operator has been inserted at some fixed position, and indeed this amplitude can be obtained from the OPE of a twist-anti twist couple in the four point amplitude. 

On general ground in the context of black hole physics it is expected for the sphere amplitudes to reproduce the Bekenstein-Hawking formula, while the higher loop amplitudes should produce quantum corrections, see \cite{Susskind:1994sm}. Borrowing this result we would expect from tree level amplitudes across a Ryu-Takayanagi surface to reproduce the classical formula for the entanglement entropy of a dual CFT, while loop amplitudes should compute quantum corrections, also in accordance with \cite{Faulkner:2013ana}. 

Here we presented a way to derive the full perturbative series in order to make consistent computations either from the target space or world sheet perspective. Some results are already present in the literature for these twisted states amplitudes; we recollect them with references in Appendix \ref{appa} together with some generalizations for sphere amplitudes with $l\geq 3$.

\section{The four points sphere}\label{4psphere}

The goal of this section is to compute the contribution to the half space entanglement entropy $S(A)$ of (\ref{sette8}) from the sphere with two couples of twist and anti-twist operators. 
The sphere with a single couple is trivial because of $SL(2,\mathbb{C})$ invariance that allows us to move around the insertion points without changing the amplitude, so that we can let the twist and anti-twist operators coincide producing the identity; in fact the usual normalization fixes $S_n^2[O_k,O_{-k},I ]=1$. The torus amplitude $T_n^2$ on the other hand had been already considered in \cite{He:2014gva} for a variety of orbifolds. For this reason the first non-trivial novel contribution comes from $S_n^2[O_k,O_{-k},O_k,O_{-k}]$, for which an expression was derived in \cite{Dixon:1986qv} \footnote{ other amplitudes at order $\lambda^2$ and higher are considerably more complicated, see the appendix for further information and relevant literature. }; furthermore this amplitude explicitly characterizes our proposal as it contains intermediate twist operator insertions whose relevance for target space entanglement entropy is at the core of our discussion. 

To start let us expand the sphere amplitude, from now on referred as $S_n^2$, around $n\sim 1$ so that
\be\label{yyy} 
S(A)|_{\text{4p-sphere}}=\lim_{n\to 1}\frac{n}{n-1}(S_1^2 + (n-1)\partial_n S_n^2 + O(n-1)^2 )=
\ee
\[
=\partial_n S_n^2|_{n=1}.
\]
In fact an analogous result also trivially follows for any amplitude, including the torus $T^2$ and higher genus Riemann surfaces without operator insertions, which is in accordance with \cite{He:2014gva}. Note here the plus sign in (\ref{yyy}) which comes from the target space $1/n$ dependence, as discussed in the last section.  To compute $S(A)$ at this order we need to analytically  continue the result for $S_n^2$ at $n\sim 1$. The task is quite non-trivial because of the sum over $k=0\dots n-1$; in fact it was noted in \cite{He:2014gva} that the result for various spin particles one loop  contributions was non analytic in $n$ unless the large $n$-limit was considered, simplifying the expressions to be later on restricted at $n\sim 1$. This recipe also produced results in accordance with older literature, and for this reason it is the one we will follow here.

We borrow the expression for the four point sphere from \cite{Dixon:1986qv} (an integration over $x$ should be included at the end):
\be 
S_n^2=\sum_{k=0}^{n-1}V_{d-2} \frac{|x(1-x)|^{2k/n(1-k/n)}}{|\prescript{}{2}{F}_{1}(\frac{k}{n},1-\frac{k}{n};1;x)|^2}Z_{\text{cl}}.
\ee
In the present work we have neglected classical contributions focusing on the local quantum oscillations around the entangling region, so we will not explicitly spell out the result for $Z_{\text{cl}}$; moreover the multiplying constant $V_{d-2}$ comes from path integration over the non orbifolded dimensions. As the total result is too complicated to be summed we will limit ourselves to a small Real $x$ expansion, capturing the effect on the target space entanglement entropy of a small worldsheet entangling region along the Real axis in between $0$ and $x$ (the other region is fixed by $SL(2,\mathbb{C})$ transformation to stretch from 1 to  $\infty$). By expanding the Hypergeometric at denominator we end up with a polynomial in $x$ starting from $1$ multiplying the $|x(1-x)|^{2k/n(1-k/n)}$ factor. The result cannot be summed in $k$ explicitly but, in the large $n$ limit, the sum on $k$ can be equivalently performed as an integration  $\int_{0}^n dk$. Using this procedure the result are terms always linear in $n$. The derivative is then trivially performed: 
\be \label{perent}
S(A)|_{\text{4p-sphere}}(x)= V_{d-2} \partial_n S_n^2|_{n=1}\propto \sqrt{\frac{\pi}{2}} \frac{\text{Erfi}\left(\frac{\sqrt{\log(x)}}{\sqrt{2}}\right)}{\sqrt{x \log(x)}}+
\ee
\[
+V_{d-2}\frac{\sqrt{x}}{4 \log(x)^{
 3/2}} \left[2 \sqrt{x} \sqrt{\log(x)} - \sqrt{2 \pi} \text{Erfi}\left(\sqrt{\log(x)}/\sqrt{2}\right) (1 + \log(x))\right]+
\]
\[
+ V_{d-2}\frac{x^{3/2}}{64 \log(x)^{5/2}} \left[6 \sqrt{x} (\sqrt{\log(x)}-5)\sqrt{\log(x)} + \sqrt{2 \pi} \text{Erfi}\left(\sqrt{\log(x)}/\sqrt{2}\right) (15 + 2\log(x)-3\log(x)^2)\right] \dots
\]
where $\text{Erfi}(z)\equiv -i\frac{2}{\sqrt{\pi}}\int_0^{i z}e^{-t^2}dt$, and higher order terms  can analogously be obtained. A simple plot shows a positive, monotonically decreasing value of $S(A)|_{\text{4p-sphere}}(x)$ in $x$ and a convex shape, up to this order. The second and third term are in fact negative and concave but still monotonically decreasing. Finally the factor  $V_{d-2}$ corresponds to the volume of the entangling boundary, as expected from the area law of entanglement entropy.  

\section{Fermions}\label{section7}

The attempt to generalize to superstrings is natural. Fermion two points twist operator correlation functions were first computed using bosonization techniques in the base $\psi_k$ in \cite{Azeyanagi:2007bj}, with the result expressed in terms of theta functions and interpreted as the usual four spin structures contributions. Later on \cite{Lokhande:2015zma} noted some problems with this interpretation as the total partition function obtained by summing other all the spin structures did not in general satisfy expected properties for Renyi entropies. Only in the limit of large and small entangling surfaces, and respectively in two opposite regimes dubbed correlated (same spin structure on each sheet) and uncorrelated (generic but definite spin structure on each sheet) it was possible to obtain a satisfactory answer. We can understand the origin of the problem by constructing the analogue of (\ref{due}) for fermionic fields:
\be\label{quattro}
\psi_k=\sum_{a=1}^{n} \psi_a e^{-i 2 \pi a \frac{k}{n}}   \;\;\;\;\;\; k=-\frac{n-1}{2},\cdots,\frac{n-1}{2}.
\ee
The issue is that the spin structure is fixed for the $\psi_a$ on the corresponding sheet but, unless it is the same for all the sheets, any field $\psi_k$ will not transform with a sign under rotation around a cycle and will instead mix up with the other $\psi_k$.  In other words if $\psi_a$ has spin structures then $\psi_k$ has not, and vice versa. The correct answer was obtained in \cite{Mukhi:2017rex} at the price of evaluating the replica partition function directly on the $n$ sheeted surface instead that using twist operator techniques. This is certainly a fine result but somehow unsatisfactory from our viewpoint, as in the end we would like to be able to obtain some twist operator computation also for fermions to be matched to the target space orbifold result.

 To do so we proceed by choosing the simplest non-trivial Riemann surface to start with, which is a torus, and replicate it $n$ times with a single cut. The generic spin structure for the fields $\psi_a$ around cycles $\alpha$ can be represented as a diagonal $n\times n$ matrix $S^{\alpha}$ acting on the vector $(\psi_1,\cdots,\psi_n)$. Inverting the expression (\ref{quattro}) as $\psi_a=U^{-1}_{ak}\psi_k$ we can immediately obtain the transformation rules for $\psi_k$:
\be
\psi_a\rightarrow S^{\alpha}_{ab}\psi_b
\ee
\be
\psi_a = U^{-1}_{ak}\psi_k \rightarrow S^{\alpha}_{ab}U^{-1}_{bk}\psi_k
\ee
\be\label{cinque}
\psi_k = U_{ka}\psi_a \rightarrow U_{ka}S^{\alpha}_{ab}U^{-1}_{bk}\psi_k.
\ee
The matrices $USU^{-1}$ can be computed easily and, unless $S$ is plus or minus the identity, they are not diagonal. For $n=2,3$ they are shown in table \ref{tabella}.
\begin{table}[h]
\hspace*{-2.5cm}
\begin{tabular}{ll}
\begin{tabular}{c}
$n=2$ \vspace{+0.2cm}\\
\begin{tabular}{c|c}
 $S$ & $USU^{-1}$   \\ \hline \\ 
$\left(+,+\right)$ & $\begin{bmatrix}
 1 & 0 \\
0 & 1    
\end{bmatrix}$   \\  \\
 $\left(+,-\right)$ & $\begin{bmatrix}
 0 & -1   \\
-1 & 0    
\end{bmatrix}$   \\ \\
 $\left(-,+\right)$ & $\begin{bmatrix}
 0 & 1  \\
1 & 0    
\end{bmatrix}$   \\ \\
 $\left(-,-\right)$ & $\begin{bmatrix}
 -1 & 0  \\
0 & -1    
\end{bmatrix}$  
\end{tabular} 
\vspace{+0.5cm}\\ 
$n=3$ \vspace{+0.2cm}\\ 
\begin{tabular}{c|c}
 $S$ & $USU^{-1}$ \\ \hline \\
$\left(+,+,+\right)$ & $\begin{bmatrix}
 1 & 0 & 0 \\
0 & 1 & 0 \\
0 & 0 & 1   
\end{bmatrix}$  \\ \\
$\left(+,+,-\right)$ & $\begin{bmatrix}
 1/3 & -2/3 & -2/3 \\-2/3 &  1/3 & -2/3 \\ -2/3 & -2/3 &  1/3  
\end{bmatrix}$   
\end{tabular} 
\end{tabular} & 
\begin{tabular}{c}
$n=3$ \vspace{+0.2cm}\\
\begin{tabular}{c|c}
 $S$ & $USU^{-1}$    \\ \hline \\
$\left(+,-,+\right)$ & $\begin{bmatrix}
 1/3 &  1/3+ i/\sqrt{3} &  1/3 -i/\sqrt{3} \\  1/3 -i/\sqrt{3} & 1/3 &  1/3+ i/\sqrt{3} \\  1/3+ i/\sqrt{3} &  1/3 -i/\sqrt{3} &  1/3
\end{bmatrix}$  \\ \\
 $\left(+,-,-\right)$ & $\begin{bmatrix}
-1/3 &  -1/3+ i/\sqrt{3} & -1/3 -i/\sqrt{3} \\ -1/3 -i/\sqrt{3} & -1/3 &  -1/3+ i/\sqrt{3} \\  -1/3+ i/\sqrt{3}  &  -1/3 -i/\sqrt{3}  &  -1/3
\end{bmatrix}$ \\ \\
$\left(-,+,+\right)$ & $\begin{bmatrix}
   1/3  &   1/3 -i/\sqrt{3}  &   1/3+ i/\sqrt{3} \\  1/3+ i/\sqrt{3}  &   1/3  &   1/3 -i/\sqrt{3} \\  1/3 -i/\sqrt{3}  &   1/3+ i/\sqrt{3}  &   1/3
\end{bmatrix}$   \\ \\
 $\left(-,+,-\right)$  &  $\begin{bmatrix}
 -1/3  &  -1/3 -i/\sqrt{3}  &   -1/3+ i/\sqrt{3} \\ -1/3+ i/\sqrt{3}  &  -1/3  &  -1/3 -i/\sqrt{3} \\ -1/3 -i/\sqrt{3}  &   -1/3+ i/\sqrt{3} &  -1/3
\end{bmatrix}$   \\ \\
 $\left(-,-,+\right)$ & $\begin{bmatrix}
 -1/3  &   2/3  &   2/3 \\  2/3  &  -1/3  &   2/3 \\  2/3  &   2/3  &  -1/3
\end{bmatrix}$  \\ \\
 $\left(-,-,-\right)$ & $\begin{bmatrix}
 -1  &  0  &  0 \\ 0  &  -1  &  0 \\ 0  &  0  &  -1  
\end{bmatrix} $
\end{tabular}
\end{tabular}
\end{tabular}
\caption{Spin structures for $\psi_a$ and their $\psi_k$ counterpart}\label{tabella}
\end{table}
How do we use this result to obtain a twist operator computation? We need to require the $\psi_k$, on the torus with two twist insertions, to transform as (\ref{cinque}) around the two cycles, which is just the orbifold construction for the group generated by the  $g=USU^{-1}$ matrices \footnote{Indeed it is a group as any element is its own inverse $g^{-1}=US^{-1}U^{-1}=USU^{-1}=g$, it contains the identity, $I=UIU^{-1}$, and it is closed under multiplication $g_1g_2=US_1S_2U^{-1}=g_{12}$}. The torus two point function $Z$ will be defined as:
\be\label{sei}
Z=\sum_{g^{\alpha},g^{\beta}}\frac{Z_{g^{\alpha}g^{\beta}}}{2^n}
\ee
with $Z_{g^{\alpha}g^{\beta}}$ the two point function with $\psi_k$ picking up the multiplicative factor $g^{\alpha}$ (resp. $g^{\beta}$) around the cycle $\alpha$ (resp. $\beta$). The result then is that we can still compute twist operator $\mathbb{Z}_n$-orbifold correlation functions provided we twist by an additional $G$-orbifold as in (\ref{sei}).

The appearance of this additional orbifold group is somehow mysterious from a target space perspective and ultimately stems from the need of using the $\psi_k$ base to introduce twist operators matching the target space $\mathbb{Z}_n$-orbifold computation. We plan to further investigate this issue.

Finally a few words on the $\gamma,\beta$ ghosts, supersymmetric completion of the $c,b$ sector. As we can express them all in terms of chiral superfields $B=\beta+\theta b$ and $C=c+\theta \gamma$  these ghosts will behave as their bosonic string counterparts, that is not being affected by any twist operator insertion.

\section{Conclusions}

In this paper we have derived a perturbative series for computing entanglement entropy in string theory for an  half-space 
entangling surface by comparing, and requiring agreement, two complementary approaches: an "inverse" replica trick construction on the target space leading to orbifold backgrounds and a world sheet twisted states computation. We have discussed in detail how to constrain the natural freedom on both sides to fix uniquely the form of the answer, and all the technical subtleties involved. In particular we have solved the puzzle of \cite{Prudenziati:2016dbc} about the right ordering among gauge fixing of the world sheet metric and the replica construction, and showed a specific form for a twist operator computation also for fermions, that was an open question from \cite{Mukhi:2017rex}. 

The first and most important development of our work would be to compute, beyond the leading terms studied here and in the literature, the entanglement entropy from (\ref{sette7}) and (\ref{sette8}). The problem is that the higher order amplitudes either are not known in closed form or known but hard to compute, and the analytic continuation in $n$ of the sum over $k$ appears even more complicated. We have summarized relevant formulas in Appendix A, where we also included some generalizations.

Many other open problems remain, as well as possible applications. Among the former the main issue is to be able to go beyond the formal solution for fermions and superstring and actually derive formulas at least for the smallest values of $l$. One possible way, at least for $l=2$, is to rely on the formalism of permutation orbifolds \cite{Klemm:1990df}, to which the spin structures reduce in this specific case. I do not have however a clear idea on how to compute higher $l$ amplitudes without going back to the n-fold covering Riemann surface result \cite{Mukhi:2017rex}.  Moreover the fermion sector solution begs for a clear target space derivation. 

In this paper we have considered flat $\mathbb{R}^d$ space and only contributions from "quantum" maps satisfying the monodromy condition (\ref{dieci}), but the generalization to compact cases and considering classical winding contributions  for which $\Delta_{C_i} X_{quantum}\neq 0$ does not look that complicated. Results are already presents in the literature, \cite{Atick:1987kd} \cite{Chen:2015cna} \cite{Datta:2013hba} and \cite{Dixon:1986qv}, and the idea presented in the paper seems to go through without main modifications. Because  this does not add much to the principal discussion however, and many other technical points had to be introduced, we decided to skip it for the moment. We plan to come back to the classical contributions and their relation with our old work \cite{Prudenziati:2016dbc} in the future. 

Even if the set up is slightly different, we expect to be able to modify the present formalism and apply it to black holes, in order to have a precise guideline to compute perturbative entanglement entropy contributions across the horizon. This may be applied both to refine the usual Bekenstein-Hawking entropy formula and, perhaps, to better understand issues of the black hole information loss paradox.  

Similarly we could consider the space-time effective action reproducing string theory up to a given $\alpha'$ order and check our result against purely field theory target space computations for entanglement entropy in higher derivative gravity (see \cite{Solodukhin:2011gn} and references therein). 

In \cite{Faulkner:2013ana} a proposal for quantum bulk corrections to the holographic entanglement entropy formula was presented, and it was shown how the first order contribution actually comes from the bulk entanglement entropy across the Ryu-Takayanagi surface. By considering an half space entangling region at the boundary this translates into a half space entanglement problem in AdS background for string theory where our formalism may be applicable. We  expect quite complicated computations in this case.

\section*{Acknowledgments}
This work has been done under financial support from the Brazilian ministries MCTI and MEC

\appendix
\section{Twist operator computations}\label{appa}

In this appendix we list various results present in the literature  and some novel generalization (mainly the  arbitrary $l$ result for the sphere 2 and 4 point functions) that can be used to compute our world sheet twist operator correlations. 

We first review the common technique for computing twist operator correlators, the stress tensor method \cite{Dixon:1986qv}. Let us consider first the sphere, and later on generalize to higher genus. The insertion of a twist field $O_k(0)$ produces a phase for the complex field $X(z,\bar{z})$ turning around (and an opposite one for $\bar{X}(z,\bar{z})$):
\be\label{sette}
X(z e^{2\pi i},\bar{z}e^{-2\pi i})=e^{2\pi i k/n}X(z ,\bar{z})
\ee
so that the expansions for $\partial X(z,\bar{z})$ and its conjugates become:
\[ 
\partial X(z)\sim \sum_{m \in \mathbb{Z}} \frac{\alpha_{m-k/n}}{z^{m+1-k/n}} \;\;\;\;\;\; \bar{\partial} X(\bar{z})\sim \sum_{m \in \mathbb{Z}} \frac{\widetilde{\alpha}_{m+k/n}}{\bar{z}^{m+1+k/n}} 
\]
\be \partial\bar{X}(z)\sim \sum_{m \in \mathbb{Z}} \frac{\bar{\alpha}_{m+k/n}}{z^{m+1+k/n}}\;\;\;\;\;\;\bar{\partial} \bar{X}(\bar{z})\sim \sum_{m \in \mathbb{Z}} \frac{\bar{\widetilde{\alpha}}_{m-k/n}}{\bar{z}^{m+1-k/n}}.
\ee
In this way the first excited states are generated by
\[
\alpha_{m-k/n} \ket{0}\neq 0\;\;\; m\leq 0, \;\;\;\;\;\; \widetilde{\alpha}_{m-k/n}\ket{0}\neq 0 \;\;\; m\leq -1
\]
\[
\bar{\alpha}_{m-k/n}\ket{0}\neq 0 \;\;\; m\leq -1, \;\;\;\;\;\; \bar{\widetilde{\alpha}}_{m-k/n}\ket{0}\neq 0 \;\;\; m\leq 0
\] 
so that the OPEs of $\partial X(z,\bar{z})$ and conjugates with  $O_k(0)$ have singular coefficients as:
\[
\partial X(z,\bar{z})\sim z^{-1+k/n} \;\;\;\;\;\;\bar{\partial} X(z,\bar{z})\sim \bar{z}^{-k/n}
\]
\[
\partial \bar{X}(z,\bar{z})\sim z^{-k/n} \;\;\;\;\;\;\bar{\partial} \bar{X}(z,\bar{z})\sim \bar{z}^{-1+k/n}.
\]

As there is not a simple operator form for the $O_k$, correlation functions of these are computed indirectly: first determine the Green function $g(z,w,z_i,w_j)$
\be\label{green}
g(z,w,z_i,w_j)\equiv\frac{-\frac{1}{2}\braket{ \partial_z X(z) \partial_w \bar{X}(w) O_k(z_1)O_{-k}(w_1)\cdots O_k(z_l)O_{-k}(w_l) }}{\braket{  O_k(z_1)O_{-k}(w_1)\cdots O_k(z_l)O_{-k}(w_l) }},
\ee
then by normal ordering ($\alpha'=2$)
\be\label{nove}
\lim_{w\rightarrow z}\left(-1/2 \;\partial_z X(z) \partial_w \bar{X}(w)-1/(z-w)^2 \right)= T(z)
\ee
we have:
\be \label{sedici}
\frac{\braket{ T(z) O_k(z_1)O_{-k}(w_1)\cdots O_k(z_l)O_{-k}(w_l) }}{\braket{  O_k(z_1)O_{-k}(w_1)\cdots O_k(z_l)O_{-k}(w_l) }} = \lim_{w\rightarrow z}\left(g(z,w,z_i,w_j)-1/(z-w)^2\right).
\ee
Finally we also know the OPE of $T$ with a primary field $O$
\be \label{otto}
T(z)O(z_i,\bar{z}_i)\sim \frac{h O(z_i,\bar{z}_i)}{(z-z_i)^2}+\frac{\partial_w O(z_i,\bar{z}_i)}{z-z_i}.
\ee
The receipt to compute the correlation function $Z(z_i,w_j)\equiv \braket{O_k(z_1)O_{-k}(w_1)\cdots O_k(z_l)O_{-k}(w_l)}$ will then be be as follows: first determine the Green function $g(z,w,z_i,w_j)$ from its asymptotic conditions, $z\rightarrow w$, $z\rightarrow (z_i,w_j)$ and $w\rightarrow (z_i,w_j)$, up to some constant $A(z_i,w_j)$. Then introduce the auxiliary Green function $h(\bar{z},w,z_i,w_j)$
\be\label{hh}
h(\bar{z},w,z_i,w_j)\equiv\frac{-\frac{1}{2}\braket{ \partial_{\bar{z}} X(\bar{z}) \partial_w \bar{X}(w) O_k(z_1)O_{-k}(w_1)\cdots O_k(z_l)O_{-k}(w_l) }}{\braket{  O_k(z_1)O_{-k}(w_1)\cdots O_k(z_l)O_{-k}(w_l) }}
\ee
which is determined analogously as $g(z,w)$ up to some constant $B(z_i,w_j)$. We can determine $A$ and $B$ from the requirement that, around any closed loop $C_i$ of total net twist zero, the quantum part of the variation of $X$ vanishes \footnote{if classical maps are considered the result can of course be non-zero, but here we will content ourselves with the purely quantum contribution.}
\be \label{dieci}
\Delta_{C_i} X_{quantum}\sim \oint_{C_i}dz g(z,w) + \oint_{C_i}d\bar{z} h(\bar{z},w)=0.
\ee
Once this is achieved we use (\ref{otto}) to produce $2l-3$ differential equations for $Z(z_i,w_j)$ in the coordinates for the unfixed operator positions on the sphere, let's say $w_2,z_3 \cdots w_l$. Let us see now in detail the procedure.

The two point function ($l=1$ with the third fixed operator being the identity) and the more complicated four point amplitude ($l=2$) were considered explicitly in \cite{Dixon:1986qv}, and we will recover them as  special cases of our more general discussion here. 

The first goal is to find the general form for the Green function  $g(z,w,z_i,w_j)$, which is determined by the local behaviour of $z$ and $w$ approaching each other and the operator positions $z_i$ and $w_j$. According to (\ref{nove}) and (\ref{sette}) they should be
\begin{align} \label{tredici}
g(z,w,z_i,w_j)&\underset{z\rightarrow w}{\sim}\frac{1}{(z-w)^2} +\; \text{finite terms} \nonumber \\
&\underset{z\rightarrow w_j}{\sim} (z-w_j)^{-k/n} \nonumber \\
&\underset{z\rightarrow z_i}{\sim} (z-z_i)^{-1+k/n}  \nonumber \\
&\underset{w\rightarrow w_j}{\sim} (w-w_j)^{-1+k/n} \nonumber \\
&\underset{w\rightarrow z_i}{\sim} (w-z_i)^{-k/n}.
\end{align}
We write the Green function as:
\be
g(z,w,z_i,w_j)=w_k(z,z_i,w_j)w_{n-k}(w,z_i,w_j)\frac{c(z,w,z_i,w_j)}{(z-w)^2}
\ee
with the cut differentials $w_k(z)$ and $w_{n-k}(w)$ accounting for the behaviour near the insertion points, which means:
\begin{align*}
w_k(z,z_i,w_j)&=\left((z-z_1)\cdots(z-z_l)\right)^{-1+k/n}\left((z-w_1)\cdots(z-w_l)\right)^{-k/n}\\
w_{n-k}(z,z_i,w_j)&=\left((w-z_1)\cdots(w-z_l)\right)^{-k/n}\left((w-w_1)\cdots(w-w_l)\right)^{-1+k/n}.
\end{align*}
$c(z,w,z_i,w_j)$ is then required to satisfy
\begin{align} \label{dodici}
&\lim_{z\rightarrow w}c(z,w,z_i,w_j)=w_k^{-1}(w,z_i,w_j)w_{n-k}^{-1}(w,z_i,w_j)  \\ \label{quattordici}
&\partial_{z}\log c(z,w,z_i,w_j)|_{z=w}+\partial_{z}\log w_k(z,w,z_i,w_j)|_{z=w}=0
\end{align} 
so we generically constrain its form to be:
\be\label{undici}
c(z,w,z_i,w_j)=A(z_i,w_j,\bar{z}_i,\bar{w}_j)(z-w)^2+C(z,w,z_i,w_j)
\ee
with 
\be 
C(z,w,z_i,w_j)=\sum_{r,s=0}^l a_{r,s}\hspace{-0.5cm}\sum_{\substack{\text{$1\leq k_1<\cdots<k_r\leq l$}\\ \text{$1\leq l_1<\cdots<l_s\leq l$}}}\hspace{-0.6cm}(z-w_{k_1})\cdots(z-w_{k_r})(z-z_{l_1})\cdots(z-z_{l_s})\cdot
\ee
\[
\cdot\hspace{-1.4cm}\prod_{\substack{\text{$i,j=1$}\\ \text{$i,j\neq k_1,\cdots ,k_r,l_1,\cdots ,l_s$}} }^l\hspace{-1.2cm}(w-z_i)\cdots(w-w_j).
\]
The coefficients $a_{r,s}$ may be functions of the $z_i, w_j$ insertion points as $A$ itself, but for notational simplicity we will momentarily keep this dependence hidden. Some symmetries reduce the total number of independent coefficients $a_{r,s}$, in particular the interchange of any $z_i \leftrightarrow z_{j}$ and/or $w_j \leftrightarrow w_{k}$ makes the $a_{r,s}$ independent of the specific values of $k_1, \cdots ,k_r$ and $l_1, \cdots ,l_s$, as already included in (\ref{undici}). Then we have symmetry under both $z_m \leftrightarrow w_m $ and $k/n \leftrightarrow 1-k/n$, and symmetry under both $z \leftrightarrow w$ and $k/n \leftrightarrow 1-k/n$. As the $a_{r,s}$ do depend also on the fraction $k/n$ we get these equivalences
\begin{align}\label{quindici}
&a_{r,s}(k/n) \hspace{+2.7cm} \leftrightarrow &a_{s,r}(1-k/n) \nonumber \\
&\updownarrow  &\updownarrow &\\ \nonumber
&a_{l-r,l-s}(1-k/n) \hspace{+1.5cm} \leftrightarrow &a_{l-s,l-r}(k/n).
\end{align}
Note in particular that the above symmetries imply $a_{m,m}=0$ as it would need to be independent of $k/n$.
We can solve equations (\ref{dodici})-(\ref{quattordici}) to determine two of the remaining inequivalent $(l+1)^2/4$ (if $l$ is odd) or $l/2(l/2+1)$ (if $l$ is even) $a_{r,s}$ coefficients to obtain $c(z,w,z_i,w_j)$ as a function of the residual $(l+1)^2/4-2$  (if $l$ is odd) or $l/2(l/2+1)-2$ (if $l$ is even) inequivalent $a_{r,s}$ and $A$. Analogously we can fix the form for the auxiliary Green function $h(\bar{z},w,z_i,w_j)$ from its singularity behaviour, which resembles (\ref{tredici}) but it is easier as it lacks the double pole term. So
\be
h(\bar{z},w,z_i,w_j)=\bar{w}_{n-k}(\bar{z},z_i,w_j)w_{n-k}(w,z_i,w_j)B(z_i,w_j,\bar{z}_i,\bar{w}_j).
\ee
The constants (in $z$ and $w$) $A$ and $B$ and the unfixed inequivalent $a_{r,s}$ are then determined (partially for $l\geq 5$) by the monodromy conditions (\ref{dieci}). To fix a basis for the loops $C_i$ we can follow the argument in \cite{Dixon:1986qv} that restricts such loops to a single sheet of the $n$-sheets cover of the original sphere with $n$th root branch cuts, as the integrals in  (\ref{dieci}) just get shifted by an overall phase as they move from one sheet to the other. We can then show on any such sheet, by cutting and gluing any possible cycle with net zero twist number, that these are indeed generated by a base of $2(l-1)$ cycles that go around the insertion points as follows
\[
C_m (w_m,z_m)\; \text{around}\; (w_m,z_m),\;\; C_{l-1+m}\; \text{around}\; (w_{l-2+m},z_{l-1+m})\;\; m=1,\cdots,l-1 .
\]

Let us show some examples of the equations determining $a_{r,s}$ for $l=1,2,3$. The first two cases were already covered in \cite{Dixon:1986qv}, the third is, to our knowledge, a novelty but for the general discussion in \cite{Knizhnik:1987xp}; also the present discussion is made in such a way to be generalizable for generic $l$ \footnote{The attentive reader may have noticed that for $l\geq 5$ we have more unfixed parameters than monodromy equations, one for $l=5$, two for $l=6$ and $l=7$ etc... I do not know at the moment how to fix this additional freedom in the Green function or how to show that it is irrelevant in the final equations for $Z(z_i,w_j)$. } (although complicated to explicitly solve for $Z(z_i,w_j)$). For $l=1$ the equations are trivial giving respectively
\[
a_{1,0}+a_{0,1}=1
\]
\[
a_{1,0}=\frac{k}{n} \;\;\;\;\;\;\; a_{0,1}=1-\frac{k}{n}
\]
whose solution also satisfies the symmetry (\ref{quindici}). As we have just two insertion points on a sphere, the third being the location of the identity operator, $SL(2,\mathbb{C})$ invariance permits to fix their value to, say, $0$ and $\infty$, producing the Green function found in \cite{Dixon:1986qv}:
\be \label{venti}
g(z,w)=w_k(z)w_{n-k}(w)\left(\left(1-\frac{k}{n}\right)\frac{z}{(z-w)^ 2}+\frac{k}{n}\frac{w}{(z-w)^ 2}  \right)
\ee
with
\[
w_k(z)=z^{-1+k/n} \;\;\;\;\; w_{n-k}(w)=w^{-k/n}
\]
As there are not independent coordinates $z_i,w_j$ the correlation function $Z(z_i,w_j)$ is a constant (fixed to 1).

More interesting is the case $l=2$ where the two equations (\ref{dodici})-(\ref{quattordici}) become:
\[
1=4 a_{1,0} + 4 a_{0,1} + a_{2,0} + a_{0,2}
\]
\[
\frac{k}{n}=3 a_{1,0}+a_{0,1}+ a_{2,0}  \;\;\;\;\;\;\; 1-\frac{k}{n}= a_{1,0}+3 a_{0,1}+ a_{0,2}
\]
whose solution respecting the symmetries (\ref{quindici}) is $ a_{2,0}=k/n$, $ a_{0,2}=1-k/n$, $a_{1,0}=a_{0,1}=0$. This again reproduces the result of \cite{Dixon:1986qv}:
\[
g(z,w)=w_k(z)w_{n-k}(w)\Big[\left(1-\frac{k}{n}\right)\frac{(w-w_1)(w-w_2)(z-z_1)(z-z_2)}{(z-w)^ 2}+
\]
\be \label{diciannove}
+\frac{k}{n}\frac{(w-z_1)(w-z_2)(z-w_1)(z-w_2)}{(z-w)^ 2} + A (z-w)^ 2 \Big]
\ee
We can then send three points to fixed values $(w_1, w_2, z_2)=(0, 1, \infty)$ leaving $z_1\equiv x$ free.

Finally for $l=3$ we have
\[
1=6 a_{1,0}+6a_{0,1}+6 a_{2,0}+6 a_{0,2}+a_{3,0}+ a_{0,3}+9a_{2,1}+9 a_{1,2}
\]  
\[
k/n=4 a_{1,0}+2a_{0,1}+5 a_{2,0}+ a_{0,2}+a_{3,0}+6a_{2,1}+3 a_{1,2}
\] 
\[
1-k/n=2 a_{1,0}+4 a_{0,1}+ a_{2,0}+5 a_{0,2}+a_{0,3}+3a_{2,1}+6 a_{1,2}.
\] 
Here we have too many independent variables and too few equations; we can express $a_{1,0}=a_1\cdot k/n$,  $a_{2,0}=a_2 \cdot k/n$, $a_{3,0}=a_3 \cdot k/n$,  $a_{2,1}=\tilde{a}_3 \cdot k/n$ and $a_{0,1},a_{0,2},a_{0,3},a_{1,2}$ with the same proportionality constants but $k/n\leftrightarrow 1-k/n$.
Solving for these constants we obtain the relationships
\[
a_1=\frac{1}{6}(a_3-1)-\frac{3}{2}\tilde{a}_3
\]
\[
a_2=\frac{1}{3}(1-a_3).
\]
determining $C(z,w,z_i,w_j)$ in (\ref{undici}) up to the numbers $a_3$ and $\tilde{a}_3$. We can easily check that the resulting Green function has the correct poles (\ref{tredici}) and it would reduce to the most straightforward generalization of (\ref{diciannove}) if we could fix $a_3=1$ and $\tilde{a}_3=0$. This however is not the correct solution of (\ref{dieci}), as we will see.

Using (\ref{sedici}) we can recover the normalized expectation value of the energy momentum tensor with twist operators, by subtracting the double pole from the Green function $g(z,w,z_i,w_j)$. For $l=2$ the result depends on $A(z_i,w_j,\bar{z}_i,\bar{w}_j)$ and it is:
\be
\frac{\braket{T(z)O_k(z_1)\cdots O_{-k}(w_2)}}{\braket{O_k(z_1)\cdots O_{-k}(w_2)}}=
\ee
\[
=\frac{1}{2}\frac{k}{n}\left(1-\frac{k}{n}\right)\left(\frac{1}{z-w_1}+ \frac{1}{z-w_2}-\frac{1}{z-z_1}-\frac{1}{z-z_2}\right)^2+\frac{A}{(z-w_1)(z-w_2)(z-z_1)(z-z_2)}.
\]
For $l=3$ instead the result is more complicated and it will not be listed explicitly here. However we do not really need the full explicit expression for our purposes but only the coefficient of the first order pole nearby the twist operator positions. This because using (\ref{otto}) with $O(z_i)$ a twist operator $O_k(z_i)$ (resp. anti twist $O_{-k}(w_j)$), and discarding the second order poles, we immediately produce differential equations in $z_i$ (resp. $w_j$) for $Z(z_i,w_j)$. In fact only $2l-3$ of these positions are free so that the counting gives one equation for $l=2$ and three for $l=3$. Having $(w_2, w_3, z_3)=(0, 1, \infty)$, renaming $(z_2, w_1, z_1)=(x_1, x_2, x_3)$ \footnote{the apparently weird conventions on the indexes are not a typo but a choice to make the comparison with the notation of \cite{Dixon:1986qv} easier } and rescaling $A\rightarrow -z_3 A$ (with $z_3\rightarrow \infty$) the three equations in $x_1,x_2$ and $x_3$ read:
\be \label{e1}
\partial_{x_1}\log Z = 
 \frac{k}{n} (1 -\frac{k}{n}) \left(\frac{1}{1 - x_1} - \frac{1}{x_1} - \frac{1}{x_1 - x_2} + \frac{1}{x_1 - x_3}\right) + 
 \ee
 \[ 
- \frac{A(x_1, x_2, x_3)}{x_1 (1 - x_1) (x_1 - x_2) (x_1 - x_3) } + \frac{(a_3-1)}{6}\frac{1}{x_1 - x_3}+ 
\]
\[
+ \tilde{a}_3 \left(-\frac{1}{ x_1} +  \frac{1}{1 - x_1}  - \frac{1}{x_1 - x_2} - \frac{3}{2}  \frac{1}{x_1 - x_3}\right)
\]
\be \label{e2}
\partial_{x_2}\log Z = 
 \frac{k}{n} (1 -\frac{k}{n}) \left(-\frac{1}{1 - x_2} + \frac{1}{x_2} + \frac{1}{x_1 - x_2} - \frac{1}{x_2 - x_3}\right) + 
 \ee
 \[ 
+ \frac{A(x_1, x_2, x_3)}{x_2 (1 - x_2) (x_1 - x_2) (x_2 - x_3) } +\frac{(a_3-1)}{6}\left(\frac{1}{x_2}-\frac{1}{1 - x_2}\right)  + 
\]
\[
+\tilde{a}_3 \left(-\frac{3}{2}\frac{1}{ x_2} +  \frac{3}{2}\frac{1}{1 - x_2}  + \frac{1}{x_1 - x_2} -  \frac{1}{x_2 - x_3}\right)
\]
\be \label{e3}
\partial_{x_3}\log Z = 
 \frac{k}{n} (1 -\frac{k}{n}) \left(\frac{1}{1 - x_3} - \frac{1}{x_3} + \frac{1}{x_2 - x_3} - \frac{1}{x_1 - x_3}\right) + 
 \ee
 \[ 
- \frac{A(x_1, x_2, x_3)}{x_3 (1 - x_3) (x_2 - x_3) (x_1 - x_3) } - \frac{(a_3-1)}{6}\frac{1}{x_1 - x_3}+ 
\]
\[
+ \tilde{a}_3 \left(-\frac{1}{ x_3} +  \frac{1}{1 - x_3}  + \frac{1}{x_2 - x_3} + \frac{3}{2}  \frac{1}{x_1 - x_3}\right).
\]
The above generalize the single differential equation in $x_1=x$ for $l=2$ that reads
\be \label{diciassette}
\partial_{x_1}\log Z = 
 \frac{k}{n} (1 -\frac{k}{n}) \left(\frac{1}{1 - x} - \frac{1}{x}\right) - \frac{A(x)}{x(1 - x) }.
 \ee
This same procedure could have been applied to $l=4$ and higher, although the expressions become increasingly more complicated.  

We move to the monodromy condition (\ref{dieci}). We have seen how to construct a basis for the cycles $C_i$, so the problem is evaluating the integrals there in order to construct equations to determine the  unfixed $A,B$ and $a_{r,s}$. After dividing by a common $w_{n-k}(w)$ the monodromy conditions becomes (the $A$ here and in the following formulas is always the rescaled function $A\rightarrow -z_3 A$ already introduced. Similarly we rescale $B\rightarrow (-z_3)^{2k/n} B$. Doing so a common factor of $(-z_3)^{k/n}$ can be pulled of from all the terms in the equation below leaving only finite pieces in the limit $z_3\rightarrow \infty$)
\be \label{diciotto}
A\oint_{C_i}dz w_k(z) + B\oint_{C_i}d\bar{z} \bar{w}_{n-k}(\bar{z})=-\oint_{C_i}dz w_k(z)\frac{C(z,w,z_i,w_j)}{(z-w)^2}.
\ee
For $l=2$ the best strategy is to send $w\rightarrow \infty$ to simplify the expression and express the integrals in terms of hypergeometric functions. From the two equations an expression in $A$ is easily achieved and from there by solving (\ref{diciassette}) the correlation function $Z(z_i,w_j)$ is obtained. The explicit expressions are available in \cite{Dixon:1986qv} and will not be repeated here. 

Instead we will attack the considerably more involved problem of $l=3$. The integral in the first term of (\ref{diciotto}) of $w_k(z)$ on $C_i$ can be evaluated in analogous way to the $l=2$ case. However, instead of an hypergeometric function we will express the result using a generalized hypergeometric of Lauricella type, whose integral representation for three variables is ($\text{Re}(c)> \text{Re}(a)>0$)
\[
F^3(b,c,a_1,a_2,a_3;y_1,y_2,y_3)=
\]
\be
=\frac{ \Gamma(c)}{\Gamma(b) \Gamma(c - b)}\int_0^1 dy \;(1 - y)^{c - b - 1} y^{
 b - 1} (1 - y_1\; y)^{-a_1} (1 - y_2\; y)^{-a_2} (1 - y_3\; y)^{-a_3}.
\ee
The integrals multiplying $A$ for the four cycles $C_i$ are listed here:
\begin{align*}
\oint_{C_1}\hspace{-0.2cm}w_k(z) =  &\;2\pi i (-1+x_2)^{-\frac{k}{n}} x_2^{-\frac{k}{n}}(-x_1+x_2)^{-1+\frac{k}{n}}\cdot \\ & \cdot F^3(1 - \frac{k}{n}, 1, \frac{k}{n},\frac{k}{n}, 1 - \frac{k}{n}, \frac{x_2-x_3}{-1+x_2}, \frac{x_2-x_3}{x_2}, \frac{x_2-x_3}{-x_1+x_2})\\
\oint_{C_2}\hspace{-0.2cm}w_k(z) = &\;2\pi i (-1)^{-\frac{k}{n}} (-x_2)^{-\frac{k}{n}} (-x_3)^{-1 +\frac{k}{n}}F^3(1 - \frac{k}{n}, 1, \frac{k}{n},\frac{k}{n}, 1 - \frac{k}{n}, x_1, \frac{x_1}{x_2}, \frac{x_1}{x_3})\\
\oint_{C_3}\hspace{-0.2cm}w_k(z) =  &\;2\pi i (-1)^{-\frac{k}{n}} (-x_1)^{-1 +\frac{k}{n}} (-x_2)^{-\frac{k}{n}}F^3(1 - \frac{k}{n}, 1, \frac{k}{n},\frac{k}{n}, 1 - \frac{k}{n}, x_3, \frac{x_3}{x_2}, \frac{x_3}{x_1})\\
\oint_{C_4}\hspace{-0.2cm}w_k(z) =  &\;2\pi i (1-x_2)^{-\frac{k}{n}} (1-x_3)^{-1 +\frac{k}{n}}F^3(1 - \frac{k}{n}, 1, \frac{k}{n},\frac{k}{n}, 1 - \frac{k}{n},1-x_1, \frac{1-x_1}{1-x_2}, \frac{1-x_1}{1-x_3}).
\end{align*}
Analogous results can be derived for the term proportional to $B$:
\begin{align*}
\oint_{C_1}\hspace{-0.2cm}\bar{w}_{n-k}(\bar{z}) =  &-2\pi i (-1+\bar{x}_2)^{-1+\frac{k}{n}} \bar{x}_2^{-1+\frac{k}{n}}(-\bar{x}_1+\bar{x}_2)^{-\frac{k}{n}}\cdot \\
& \cdot F^3(\frac{k}{n}, 1,1- \frac{k}{n},1-\frac{k}{n}, \frac{k}{n}, \frac{\bar{x}_2-\bar{x}_3}{-1+\bar{x}_2}, \frac{\bar{x}_2-\bar{x}_3}{\bar{x}_2}, \frac{\bar{x}_2-\bar{x}_3}{-\bar{x}_1+\bar{x}_2})\\
\oint_{C_2}\hspace{-0.2cm}\bar{w}_{n-k}(\bar{z})= &\;2\pi i (-1)^{\frac{k}{n}} (-\bar{x}_2)^{-1+\frac{k}{n}} (-\bar{x}_3)^{-\frac{k}{n}}F^3(\frac{k}{n}, 1,1- \frac{k}{n},1-\frac{k}{n}, \frac{k}{n}, \bar{x}_1, \frac{\bar{x}_1}{\bar{x}_2}, \frac{\bar{x}_1}{\bar{x}_3})\\
\oint_{C_3}\hspace{-0.2cm}\bar{w}_{n-k}(\bar{z}) =  &\;2\pi i (-1)^{\frac{k}{n}} (-\bar{x}_1)^{-\frac{k}{n}} (-\bar{x}_2)^{-1+\frac{k}{n}}F^3(\frac{k}{n}, 1,1- \frac{k}{n},1-\frac{k}{n}, \frac{k}{n}, \bar{x}_3, \frac{\bar{x}_3}{\bar{x}_2}, \frac{\bar{x}_3}{\bar{x}_1})\\
\oint_{C_4}\hspace{-0.2cm}\bar{w}_{n-k}(\bar{z})=  &-2\pi i (1-\bar{x}_2)^{-1+\frac{k}{n}} (1-\bar{x}_3)^{- \frac{k}{n}}F^3(\frac{k}{n}, 1,1- \frac{k}{n},1-\frac{k}{n}, \frac{k}{n},1-\bar{x}_1, \frac{1-\bar{x}_1}{1-\bar{x}_2}, \frac{1-\bar{x}_1}{1-\bar{x}_3}).
\end{align*}
The first term on the right of (\ref{diciotto}) is more complicated. The first step is to choose a convenient limit for $w$ as the final result should be independent of it. For $l=2$ the most convenient choice was $w\rightarrow\infty$, but in this case we found the integral to diverge. Instead we pick $w\rightarrow z_2=x_1$ for the integrals on cycles $C_1$ and $C_3$ while $w\rightarrow z_1=x_3$ for the integrals on cycles $C_2$ and $C_4$. This term depends on the various coefficients $a_{r,s}$, themselves functions of the two constants $a_3,\tilde{a}_3$; simplifying somehow the expressions we end up with an integrand for cycles $C_{1,3}$ proportional  to \footnote{ stripped of a factor going as $\sim -z_3$ that together with a term $(z-z_3)^{-1+k/n}$ inside  $w_k(z)$ produces the overall $-z_3^{k/n}$ as in the other two terms of (\ref{diciotto}) }:
\[
\frac{C(z,w\rightarrow x_1)}{(z-x_1)^2}=(1 - \frac{k}{n})\frac{ (x_1-1) x_1 (x_1 - x_2)  (z-x_3)}{z-x_1 } +
 \frac{1}{6} \frac{1 - a_3}{z-x_1} \Big((1 - \frac{k}{n}) (x_1-1) x_1 (x_1 - x_2)\cdot 
 \]
 \[\cdot \big((x_1 - x_3) -4 (z-x_3) \big) + (z-1) z (z-x_2) (z-x_3) \Big(-\frac{k}{n} \frac{z-x_1}{z-x_3} + (1 - \frac{k}{n}) \big(2 \frac{(x_1-1) x_1}{(z-1) z} + 
\]
\[
+2 \frac{(x_1-1) (x_1 - x_2)}{(z-1) (z-x_2)} + 2 \frac{x_1 (x_1 - x_2)}{z (z-x_2)} - \frac{x_1-1}{z-1} - \frac{x_1}{z} - \frac{x_1 - x_2}{z-x_2 }\big)\Big)\Big)+ 
\]
\[+ \frac{\tilde{a}_3}{z-x_1} \Big(-\frac{3}{2} (z-1) z (z-x_2) (z-x_3) \big((1 - \frac{k}{n}) (\frac{ x_1-1}{z-1} + \frac{x_1}{z} + \frac{x_1 - x_2}{z-x_2}\big) + \frac{k}{n} (1 + \frac{x_1 - x_3}{z-x_3})\big) + 
\]
\[
+(x_1-1) x_1 (x_1 - x_2) \big((1 - k/n) ((z-x_3) \frac{z-1}{x_1-1} + \frac{z}{x_1} + \frac{z-x_2}{x_1 - x_2}) + (x_1 - x_3) ((\frac{z-1}{x_1-1} + \frac{z}{x_1} + \frac{z-x_2}{x_1 - x_2}) - 3/2)\big)+
\]
\[ + (x_1 - x_3) \frac{k}{n} (\frac{(z-1) z}{(x_1-1) x_1} + \frac{(z-1) (z-x_2)}{(x_1-1) (x_1 - x_2)} + \frac{z (z-x_2)}{x_1 (x_1 - x_2)})\Big)\Big)
\]
A similar expression can be found for the cycles $C_{2,4}$ for $C(z,w\rightarrow x_3)/(z-x_3)^2$. Together with $w_k(z)$ these integrands lead to generalized Lauricella hypergeometric functions in a similar fashion as the other terms in $A$ and $B$, but with more complicated expressions. 

If for a moment we fix $a_3=1,\;\tilde{a}_3=0$ many simplifications occur in the above expression for $C(z,w\rightarrow x_1)/(z-x_1)^2$ (and the corresponding one in $x_3$) so that, by carefully massaging the hypergeometrics with Euler and Pfaff transformations (see for example (3.30) of \cite{SHY-DER-LIN}), we can solve in $A$ the monodromy conditions for, say, cycles $C_1, C_3$. In analogy with the $l=2$ case we can plug this $A$ function in equation (\ref{e1}), also considerably simplified by this assumption, and solve for $Z$. Explicitly (reiterating, this is valid if $a_3=1,\;\tilde{a}_3=0$)
\[
A(x_1,x_2,x_3)\overset{?}{=} (1-x_1)x_1(x_1-x_2)(x_1-x_3)\partial_{x_1}\log f(x_1,x_2,x_3)
\]
with
\[
f(x_1,x_2,x_3)=\Big(F^3(\frac{k}{n}, 1, -\frac{k}{n},\frac{k}{n}, 1 - \frac{k}{n},1-\frac{x_3}{x_2}, \frac{x_3-x_2}{x_2(x_3-1)}, \frac{x_1(x_2-x_3)}{x_2(x_1-x_3)}) \cdot 
\]
\[
\cdot F^3(1-\frac{k}{n}, 1, -1+\frac{k}{n},1-\frac{k}{n},\frac{k}{n},\frac{\bar{x}_3}{\bar{x}_2}, \frac{\bar{x}_3(\bar{x}_2-1)}{\bar{x}_2(\bar{x}_3-1)}, \frac{\bar{x}_3(\bar{x}_1-\bar{x}_2)}{\bar{x}_2(\bar{x}_1-\bar{x}_3)}) +
\]
\[
+F^3(\frac{k}{n}, 1, -\frac{k}{n},\frac{k}{n}, 1 - \frac{k}{n},\frac{x_3}{x_2}, \frac{x_3(x_2-1)}{x_2(x_3-1)}, \frac{x_3(x_1-x_2)}{x_2(x_1-x_3)})\cdot\]
\[
\cdot  F^3(1-\frac{k}{n}, 1, -1+\frac{k}{n},1-\frac{k}{n}, \frac{k}{n},1-\frac{\bar{x}_3}{\bar{x}_2}, \frac{\bar{x}_3-\bar{x}_2}{\bar{x}_2(\bar{x}_3-1)}, \frac{\bar{x}_1(\bar{x}_2-\bar{x}_3)}{\bar{x}_2(\bar{x}_1-\bar{x}_3)}) \Big)
\]
and plugging inside equation (\ref{e1}) we would get
\[
Z\overset{?}{=}\text{const} |(1-x_1)x_1(x_1-x_2)(x_1-x_3)|^{-2\frac{k}{n}(1-\frac{k}{n})}f(x_1,x_2,x_3).
\]
The problem of this assumption is that an expression for $A$ can also be obtained from the monodromy condition for cycles $C_2, C_4$, and it is different from the one above, so leading to an inconsistency. The correct solution then passes through determining $a_3$ and $\tilde{a}_3$ in such a way that the two expressions agree, but then we should face the considerably more complicated problem of solving the full equations  (\ref{e1},\ref{e2},\ref{e3}) with this cumbersome $A$ function inside. 

I choose not to write down the final expression for $A$ (and $B$ which is not necessary), $a_3$ and $\tilde{a}_3$ as they are quite intricate functions of many hypergeometrics that I could not put in a usable form to ultimately integrate equations  (\ref{e1},\ref{e2},\ref{e3}). I can provide the explicit expressions upon request.

Higher genus amplitudes of twist operators were computed in \cite{Atick:1987kd}. Formulas are explicit but contain complicated integrals that can only be performed in certain limits. We will not review the full work but instead focus on the modifications that need to be implemented to apply their results to the present contest. In \cite{Atick:1987kd} the steps reviewed in this appendix for constructing the sphere amplitudes are modified for a generic genus $g$ surface $\Sigma_g$. Green functions are defined as in (\ref{green}) and (\ref{hh})  but now on $\Sigma_g$, and their local properties are the same as (\ref{tredici}). However the monodromy condition (\ref{dieci}) is now richer as new closed loops can be found in comparison with the sphere. In particular we illustrate the procedure for the torus $T^2$ but it can be, at least formally, extended to higher genus; in this case we have two more closed cycles under which the green function should be periodic: $z\rightarrow z+1$ and $z\rightarrow z+\tau$. As usual we place the twist operators in $z_i$ and the anti-twist operators in $w_j$, $l+l$ of them in total. The cut differentials $w_k, w_{n-k}$ are now modified to be of the form:
\be 
w_k^{\alpha_k}(z,z_i,w_j,z_{\alpha_m})=\prod_{i=1}^l \Theta_1(z-z_i)^{-1+\frac{k}{n}}\prod_{j=1}^l\Theta_1(z-w_j)^{-\frac{k}{n}}\Theta_1(z-z_{\alpha_k}-Y)\hspace{-0.3cm}\prod_{m=1, \; m\neq k}^l\hspace{-0.3cm}\Theta_1(z-z_{\alpha_m})
\ee
and
\be 
w_{n-k}^{\beta_k}(z,z_i,w_j,z_{\beta_m})=\prod_{i=1}^l \Theta_1(z-z_i)^{-\frac{k}{n}}\prod_{j=1}^l\Theta_1(z-w_j)^{-1+\frac{k}{n}}\Theta_1(z-z_{\beta_k}-Z)\hspace{-0.3cm}\prod_{m=1, \; m\neq k}^l\hspace{-0.3cm}\Theta_1(z-z_{\beta_m})
\ee 
with
\be 
Y=-\sum_{m=1}^l z_{\alpha_m}+\sum_{i=1}^l(1-\frac{k}{n})z_i+\sum_{j=1}^l \frac{k}{n}w_j
\ee
and
\be 
Z=-\sum_{m=1}^l z_{\beta_m}+\sum_{i=1}^l \frac{k}{n}z_i+\sum_{j=1}^l(1-\frac{k}{n})w_j.
\ee
These cut differentials are now expressed in terms of Theta functions, but to be really periodic in $z\rightarrow z+1$ and $z\rightarrow z+\tau$ and offset the phases that would otherwise appear we multiply by $l$ additional Theta functions $\Theta_1(z-z_{\alpha_m})$ and $\Theta_1(z-z_{\beta_m})$ (that makes them periodic under $z\rightarrow z+1$) and require one $\alpha_m$ and $\beta_m$ respectively to be changed into $Y$ and $Z$ (to be periodic  under $z\rightarrow z+\tau$). This constructs $l+l$ independent $w_k^{\alpha_k}$ and $w_{n-k}^{\beta_k}$. Note that the present formalism is consistent with our requirements to have the same index $k$ for all twist operators (and $-k$ for the anti twist), whereas the
correlation functions of \cite{Atick:1987kd} were constructed out of twist operators of always positive twist $k/n$ and with the only requirement in order not to vanish that $\sum_i k_i/n=M \;(M\in \mathbb{N})$ (here $M$ does not appear and instead of $L-M$ and $M$ independent cut differential we have $l+l$ of them in a much more symmetric structure). Thus the formalism of \cite{Atick:1987kd} should be adapted to our present purposes by the substitution of the cut differentials there to the ones here. The rest of the discussion remains basically the same.

\end{document}